%% file: paper.tex
\documentclass[sigconf,10pt]{acmart}
\usepackage{caption}
\usepackage{amsmath,calc,epsfig,xspace}
\newcommand{\tightText}{
  \usepackage[compact]{titlesec}
}
\tightText
\titlespacing*{\section}
{0pt}{1.5ex}{1ex}
\titlespacing*{\subsection}
{0pt}{1ex}{0.5ex}

\ccsdesc[500]{Information systems~Information storage systems}
\ccsdesc[500]{Computer systems organization}

\keywords{Caching, Data Freshness, Time to Live (TTL), Cache Invalidation, Cache Update}
  
\copyrightyear{2024}
\acmYear{2024}
\setcopyright{rightsretained}
\acmConference[HOTNETS '24]{The 23rd ACM Workshop on Hot Topics in Networks}{November 18--19, 2024}{Irvine, CA, USA}
\acmBooktitle{The 23rd ACM Workshop on Hot Topics in Networks (HOTNETS '24), November 18--19, 2024, Irvine, CA, USA}
\acmDOI{10.1145/3696348.3696858}
\acmISBN{979-8-4007-1272-2/24/11}

\newcommand{\hide}[1] 
{
\ifthenelse{\boolean{false}}{#1}{}
}
\usepackage[T1]{fontenc}
\usepackage{textcomp}
\usepackage{setspace}
\usepackage[title]{appendix}

\usepackage{xcolor}

\usepackage{times}
\usepackage{float}
\usepackage{subcaption}
\usepackage{multirow}
\usepackage{array}
\usepackage{etoolbox}
\usepackage{breakurl}
\usepackage{makecell}

\usepackage{booktabs}
\usepackage{amsmath}
\usepackage[linesnumbered,ruled]{algorithm2e}
\usepackage[noend]{algpseudocode}
\usepackage{csquotes}
\usepackage[htt]{hyphenat}
\usepackage{upgreek}
\usepackage{url}

\definecolor{mygreen}{rgb}{0.1,0.5,0.1}
\definecolor{myblue}{rgb}{0.1,0.1,0.7}
\definecolor{mycmtcol}{rgb}{0.5,0.0,0.5}

\usepackage[hyperref]{}
\hypersetup{              
  colorlinks,
  linkcolor=myblue,
  citecolor=mygreen,
  urlcolor=myblue,
}

\newcommand{\tput}{\ensuremath{C_F}\xspace}
\newcommand{\lat}{\ensuremath{C_S}\xspace}
\newcommand{\staleness}{\lat}
\newcommand{\freshness}{\tput}

\usepackage{epstopdf}
\usepackage{graphicx}
\usepackage[T1]{fontenc}
\usepackage[utf8]{inputenc}
\usepackage{blindtext}

\usepackage{arydshln}
\usepackage[normalem]{ulem}

\newcommand{\myparagraph}[1]{\vspace{0.5em}\noindent {\bf #1}}

\newcommand\module{\@startsection{paragraph}{4}{\z@}%
                                    {1.25ex \@plus1ex \@minus.2ex}%
                                    {-1em}%
                                    {\normalfont\normalsize\slshape\bfseries}}

\usepackage{listings}
\usepackage{zi4} 

\makeatletter
\usepackage{fancyhdr}

\input{sections/title}

\begin{document}

\input{sections/abstract}

\maketitle

\input{sections/intro}

\input{sections/background}
\input{sections/design}
\input{sections/related_works}
\input{sections/design_realization}

\bibliographystyle{ACM-Reference-Format} 
\bibliography{biblio.bib}

\end{document}

%% file: sections/title.tex
\title{Revisiting Cache Freshness for \\ Emerging Real-Time Applications}

\author{
Ziming Mao$^\dagger$\;
Rishabh Iyer$^\dagger$\;
Scott Shenker $^{\dagger\ast}$\;
Ion Stoica $^\dagger$\;
\\
\emph{\normalfont{$^\dagger$UC Berkeley}}
\emph{\normalfont{$^\ast$ICSI}}
}

%% file: sections/abstract.tex
\begin{abstract}
Caching is widely used in industry to improve application performance by reducing data-access latency and taking the load off the backend infrastructure. TTLs have become the de-facto mechanism used to keep cached data reasonably fresh (i.e., not too out of date with the backend). 
However, the emergence of real-time applications requires tighter data freshness, which is impractical to achieve with TTLs.
We discuss why this is the case, and propose a simple yet effective adaptive policy to achieve the desired freshness.
\end{abstract}

%% file: sections/intro.tex
\section{Introduction}
\label{sec:introduction}

\begin{figure}[t!]{
    \centering
    \includegraphics[width=\columnwidth]{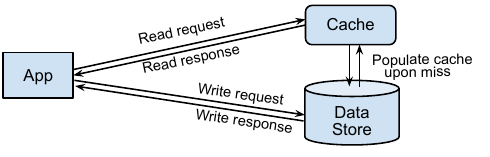}
    \caption{Lazy or cache-aside caches; the predominant way in which in-memory caches are deployed today. In such caches, data freshness is not guaranteed since writes bypass the cache. Servicing a miss can either be initiated by the cache or by the application.}
    \vspace{-1em}
    \label{fig:system-model}
}
\end{figure}

In-memory caching is widely used to improve application performance by reducing data-access latency and the load on the backend data store. 
The vast majority of these caches are deployed as lazy or cache-aside caches~\cite{sultan2024ttls, twitter-cache, aws_caching, nishtala2013scaling} (shown in \autoref{fig:system-model}). 
In such caches, reads are served from caches, writes are issued directly to the backend data store, and the caches are populated when read misses in the cache. 
 
A key aspect of caching is \emph{data freshness}. Data is fresh within a staleness bound $T$ if a cached object reflects the state of the backend data store (the ground truth) at some point in the last $T$ seconds~\cite{baseball}. 

The primary technique used to ensure data freshness in caches today is Time-To-Live (TTL)~\cite{cachelib, twitter-cache, segcache, sultan2024ttls, faascache, modeling-ttl, perf-eval-ttl, exact-ttl-analysis, adaptive-ttl, ttl-cloud-caches, cost-aware-ttl}.
TTLs are typically on the order of minutes to hours~\cite{twitter-cache}, and work as follows: 
whenever a data object is brought into the cache from the data store, a timer of duration $T$ is set.
When the timer expires, the object is either (1)~re-fetched from the data store or (2)~expired and removed from the cache; both of these actions ensure that future reads see a fresh copy from the data store. 
The main advantage of TTLs is that they are easy to deploy because they need little coordination between the cache and the data store; all freshness decisions can be made using a simple timer local to the cache. 
This is a primary reason behind the popularity of TTL-based techniques to bound staleness for over two decades~\cite{DBLP:journals/tcs/CohenHK05,DBLP:conf/infocom/JungBB03}.

However, the emergence of \emph{real-time} applications with tighter freshness requirements demands new solutions for cache freshness~\cite{adya2023pulling}. 
For example, Databricks' Unity Catalog~\cite{databricks_unity_catalog} stores metadata information that requires high data freshness, on the order of seconds. Other examples include serving dynamic web content~\cite{abolhassani2021fresh}, financial applications (e.g., viewing stock prices)~\cite{cipar2014trading, bright2002using}, 
ad bidding~\cite{ads-bidding},  emergency response~\cite{bright2002using}, and real-time monitoring and diagnosis~\cite{mao2024trinity}.
These applications have stringent freshness requirements since they typically entail real-time decision-making.
For example, a service that provides stock information to analysts requires data to be as fresh as possible to enable real-time financial decisions. A service managing Access Control List (ACL) needs to be fresh to ensure that permissions can be added or revoked immediately.

TTLs introduce prohibitive overhead when applications require data freshness at real-time timescales.
This is because, with TTLs, the rate at which additional read requests are made to the backend---either to re-fetch data or due to cache misses that occur when data is expired---is inversely proportional to $T$ since these requests are made each time the timer with duration $T$ expires.
This leads to prohibitively high overheads when $T$ is small. 
This overhead is so large that when designing systems for real-time applications, practitioners are forced to sacrifice caching (and its benefits) for data freshness, preferring to issue reads directly to the database instead.  

\vspace{0.25em}
So, we ask the question: \textit{Is it feasible to 
efficiently provide cache freshness for real-time applications}? 
\vspace{0.25em} 

To answer this question, we first develop a simple mathematical model that enables us to better understand the trade-offs presented by different techniques for ensuring cache freshness. We then use this model to show that, \emph{at real-time timescales, making freshness decisions in response to incoming writes is more efficient than TTL-based policies}. 
Based on this observation, we develop a simple algorithm that adapts to the read-write characteristics of the incoming workload, and show, using simulations, that it has the potential to answer the above question in the affirmative. A salient benefit of our algorithm is that it makes freshness decisions on a \textit{per-object} basis; this ensures that it can be implemented efficiently since it does not require coordinating states across objects.

While providing a theoretical model to reason about cache freshness, our work leaves several system-design questions unanswered. 
In particular, reacting to writes mandates active coordination between the backend and the cache, a topic that has received little attention due to the near-ubiquitous use of TTLs thus far. 
We conclude this paper with a set of open research questions that must be answered before real-time freshness can be realized in a practical system.

%% file: sections/background.tex
\section{Reasoning About Freshness Quantitatively}
\label{sec:motivation}

\begin{figure*}[t!]
\centering
\begin{subfigure}{.26\linewidth}
    \centering
 \includegraphics[width=\linewidth]{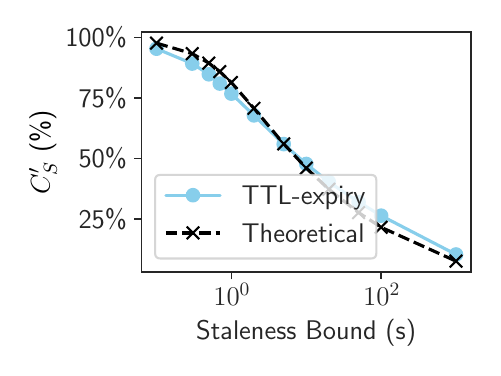}
    \vspace{-.3in}
    \caption{Poisson
    }
\end{subfigure}
~
\begin{subfigure}{.26\linewidth}
    \centering
  \includegraphics[width=\linewidth]{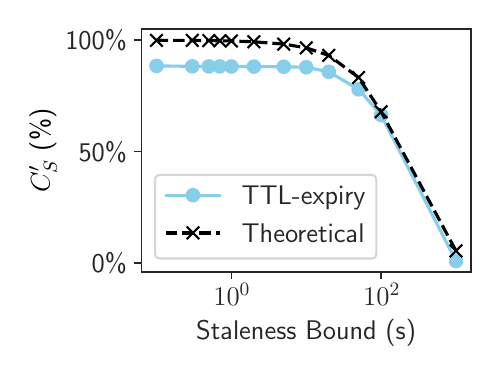}
    \vspace{-.3in}
    \caption{Meta
    }
\end{subfigure}
~
\begin{subfigure}{.26\linewidth}
    \centering
 \includegraphics[width=\linewidth]{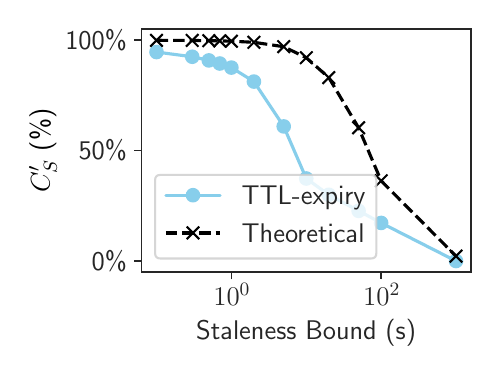}
    \vspace{-.3in}
    \caption{Twitter
    }
\end{subfigure}
\caption{Effect of decreasing staleness bound  on normalized staleness cost ($C_S'$, \S\ref{sec:motivation}). The x-axis is in log scale.}
\label{ttl-bad}
\end{figure*}

\begin{figure*}[t!]
\centering
\begin{subfigure}{.26\linewidth}
    \centering
 \includegraphics[width=\linewidth]{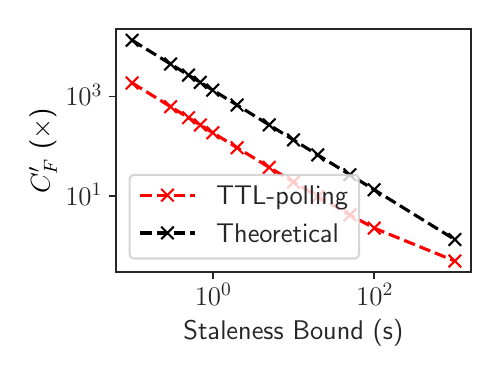}
    \vspace{-.3in}
    \caption{Poisson
    }
\end{subfigure}
~
\begin{subfigure}{.26\linewidth}
    \centering
  \includegraphics[width=\linewidth]{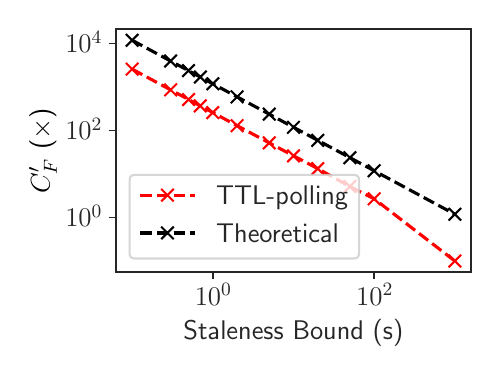}
    \vspace{-.3in}
    \caption{Meta
    }
\end{subfigure}
~
\begin{subfigure}{.26\linewidth}
    \centering
 \includegraphics[width=\linewidth]{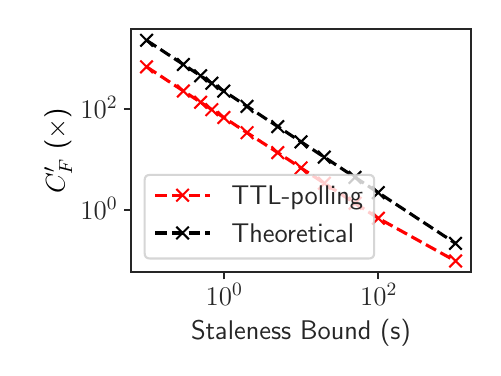}
    \vspace{-.3in}
    \caption{Twitter
    }
\end{subfigure}
\caption{Effect of decreasing staleness bound on normalized freshness cost ($C_F'$, \S\ref{sec:motivation}). Both the x and the y axis are in log scale.}
\label{polling-bad}
\end{figure*}

We now introduce a simple mathematical model that enables us to quantitatively reason about the trade-offs presented by different techniques for ensuring cache freshness (\S\ref{sec:background-model}). 
We then validate our model by showing how it can model the overheads of TTL-based policies at various timescales (\S\ref{sec:background-ttl}).

\subsection{The cost of serving fresh data}
\label{sec:background-model}
Since writes bypass caches, cached data is not guaranteed to be fresh. 
Thus, serving fresh data from the cache incurs cost (in terms of overhead on the infrastructure). 
We model this cost using two metrics: the freshness cost (\freshness), and the staleness cost (\staleness). 
\freshness refers to the \textit{throughput} overhead incurred 
to keep data fresh in the cache.
\freshness captures 
the overhead (\textit{e.g.,} compute and network) of sending and receiving messages between the cache and the data store to keep data in the cache fresh: including backend invalidating or updating data in the cache upon write, or the cache fetching fresh data from the backend when a miss occurs due to stale data.
\freshness aggregates the 
overhead across different parts of the system into a single metric\footnote{Since the policy can be implemented differently and across various parts of the system (e.g., the cache, the backend, the load balancer, etc), we chose to aggregate them into a single metric for simplicity. We elaborate on this choice in \S\ref{realizing-systems-design}. }.

The staleness cost (\staleness) refers to the \textit{latency} overhead incurred when reading data in the cache that is \emph{not} fresh (i.e., stale). 
This overhead manifests as increased end-to-end \emph{latency} for clients since stale data causes a request to miss in the cache.
As the precise latency is a function of the system implementation, we quantify \staleness in terms of the \emph{number of cache misses} that occur when the requested object was present in the cache, but could not be returned since it was stale. \staleness is different from miss ratio, which \textit{additionally} considers the misses as a result of reading un-cached data (data that was evicted or never brought into the cache). 

We use \freshness and \staleness to compare the throughput and latency overheads of different mechanisms to ensure freshness. 
To calculate \freshness and \staleness for entire workloads, we make a simplifying assumption that \staleness and \freshness for different data objects are \emph{independent}, and so \staleness and \freshness for the entire workload is the sum of \staleness and \freshness for each object accessed in the workload. This assumption does not strictly hold; for instance, \staleness is affected by whether the object is evicted from the cache (which is a function over all objects). 
However, we find that it allows for a simple formulation of \staleness and \freshness while providing results that closely match the simulations. 

\subsection{Why TTLs are no longer sufficient}
\label{sec:background-ttl}

TTLs are deployed in two forms: TTL-expiry and TTL-polling. 
In the former, when the TTL expires, the object is invalidated in the cache with the next read incurring a miss. 
In the latter, when the TTL expires, the object is re-fetched from the data store, to ensure that subsequent reads see fresh data. 

We now evaluate \staleness and \freshness for the above TTL-based policies. We use the ``bounded staleness'' definition of freshness introduced in \S\ref{sec:introduction}: cached data is considered fresh if it reflects all writes made $\geq T$ time ago to the backend data store. Since we assume that \staleness and \freshness for different objects are independent and additive, we consider each object independently. Let $P_R(T)$, $P_W(T)$ be the probability that there exists at least one read, or one write over an interval $T$ to that object. 
To calculate $P_R(T)$ and $P_W(T)$, one way is to model the request arrival as a Poisson process with an average rate $\lambda$.
Like most prior work~\cite{jung2003modeling, abolhassani2021fresh, yu1999scalable}, we assume that individual requests to the object are independent and are reads with a probability of $r$ and writes with a probability of $1-r$. 
In this case,  
$P_R(T) = 1 - e^{-\lambda r T}$ and
$P_W(T) = 1 - e^{-\lambda(1-r)T}$. 

\myparagraph{TTL-expiry:} To calculate \staleness, we consider a time period of $T'$. 
Since data is expired every $T$, the number of misses incurred is $1$ per interval $T$, if there was at least one read request during that interval. 
Therefore, 
\staleness over a time interval $T'$ is:
$C_S = \frac{T'}{T} P_R(T)$. The miss ratio due to reading cached but stale data is the staleness cost divided by the total number of reads over $T'$ 
($N_R$, $N_R = \lambda r T'$ under Poisson): $\frac{C_S}{N_R}$. As $T \rightarrow 0$, the miss ratio approaches 1. 
Since TTL-expiry does not need coordination with the backend to keep data in the cache fresh, the only overhead incurred as part of \freshness is those to service misses due to stale data. So, $C_F = C_S \times c_m$, where $c_m$ is the overhead incurred upon a miss. 

\myparagraph{TTL-polling:} For this policy, the staleness cost (\staleness) is zero. 
This is because TTL-polling proactively fetches data from the backend when the TTL expires, ensuring that any data present in the cache is never stale.
However, this leads to a large \freshness. 
Specifically, \freshness over a time $T'$ is $
C_F = c_m \times \frac{T'}{T}$. This is because, at the end of each $T$, the cache must read the fresh value from the backend data store, just as it would during a miss. Once again, as $T \rightarrow 0$, \freshness increases significantly.

To demonstrate how large these overheads can get in practice, and also as a sanity check for our simple model, we perform simulations that measure the freshness and staleness costs.
We simulate three workloads; all of which consist of multiple keys with limited cache capacity; 
to evaluate our assumption about \staleness and \freshness being additive. 
The three workloads are a synthetic Poisson workload with $\lambda=10$ and Zipfian distribution ($s=1.3$) across keys, and two production workloads from Meta~\cite{cachelib, cachebench-fb-hw-eval} and Twitter~\cite{twitter-cache}, respectively.

To give a better idea of how much these overheads matter, we normalize both \freshness and \staleness. 
\freshness is normalized ($C'_F$) by the overhead incurred to serve all read requests in the system. 
Thus $C'_F$ represents the ratio of the wasted cycles to the useful cycles spent serving data in the system. 
We normalize \staleness ($C'_S$) by the number of reads for which the object requested was present in the cache. 
Thus $C'_S$ represents the miss ratio caused solely due to reading stale data. 

\autoref{ttl-bad} and \autoref{polling-bad} illustrate the results for TTL-expiry and TTL-polling respectively\footnote{We show only $C'_S$ for expiry since $C'_F$ is a simple multiple with $c_m$. We plot only $C'_F$ for polling since $C'_S$ for polling is zero}, compared against our theoretical model. 
We see two clear takeaways: (1)~for both policies, our model predicts the overhead with reasonable accuracy, despite our assumption of \freshness and \staleness being additive and independent (2)~the overhead increases to prohibitive amounts as $T$ shrink close to $0$.
Practitioners today are aware of the latter, and as a result, sacrifice caching (and its benefits) when building systems for applications that require real-time freshness.

%% file: sections/design.tex
\section{Design}
\label{sec:design}

Our proposed approach is based on the observation that freshness decisions (e.g., whether to expire or re-fetch cached data) are only necessary when the system receives \emph{write} requests. 
Specifically, to ensure bounded staleness of $T$ for a particular object, the data store and the cache only need to coordinate once per $T$ if one or more write requests to that object were received during the past $T$, and need not coordinate otherwise.

Based on this observation, we propose an approach that reacts to writes with either \emph{updates} or \emph{invalidates}. 
An update is a message from the backend to the cache that \textit{modifies} an object in the cache to reflect its latest state, and importantly \textit{does nothing} if the object is not in the cache. 
An invalidate is a message from the backend to the cache that marks a cached object as stale or invalid, causing the following read to be treated as a miss.
Updates and invalidates are counterparts to TTL-polling and TTL-expiry; both updates and TTL-polling refresh cached objects, while invalidates and TTL-expiry expire cached objects. 
The only difference is that updates and invalidates are performed when a write occurs, while the two TTL-based policies are used when the TTL expires.

\begin{figure}[t!]{
    \centering
    \includegraphics[width=.75\columnwidth]{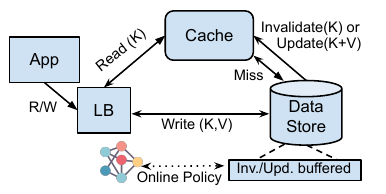}
    \caption{System Overview. Depending on the workload pattern, the policy dynamically decides between invalidation and updates. Invalidates or updates are buffered at the data store and batched over $T$.}
    \label{fig:architecture}
}
\end{figure}

\autoref{fig:architecture} shows our proposed system architecture. 
New invalidates or updates over $T$ are buffered and batched at the data store. 
Depending on the policy (which we will discuss in the rest of the section), the backend either sends out invalidates or updates for the buffered keys every interval of $T$. 
Note, that this buffering of writes and sending of updates and invalidates can also be implemented at proxies, not just at the data store. 

In the rest of the section, we first use our model to show that updates and invalidates typically lead to lower overheads than TTL-polling and TTL-expiry, respectively to make the case for reacting to writes and not TTLs (\S\ref{bounding-argument}).
We then explore the question of how to choose between sending an update or an invalidate upon receiving a write and show that different keys benefit from different decisions (\S\ref{adaptive-algorithm}). 

\subsection{Reacting to writes versus TTLs}
\label{bounding-argument}

We now calculate the freshness cost \freshness and staleness cost \staleness for updates and invalidates, and show that they are typically lower than the overheads for TTL-based policies at real-time timescales.
To do so, we introduce two additional parameters in our model $c_u$ and $c_i$, which refer to the overhead of updates and invalidates, respectively. We assume $c_u < c_m$ (i.e., it is cheaper to update than to incur a miss). 

\myparagraph{Updates are more efficient than TTL-polling.}
Consider a period $T'$. 
Our solution only requires sending one update every $T$ in case of one or more writes during that duration. 
Hence, since the probability of at least one write over $T$ is $P_W(T)$, the freshness cost of refreshing a single key over $T'$ is $C_F = \frac{T'}{T}P_W(T) \times c_u$. 
In comparison, \freshness for TTL-polling over $T'$ is $c_m \times \frac{T'}{T}$. 
Since $c_m > c_u$ and $P_W(T) < 1$, we conclude that updates have lower throughput overhead.
In terms of \staleness, both the above policies proactively keep data fresh, and so $\staleness=0$, making updates more efficient than their TTL counterparts.

\myparagraph{Invalidation is more efficient than TTL-expiry.} Consider a time period $T'$ broken into multiple intervals of duration $T$. Consider two consecutive intervals $T_{0}$ and $T_{1}$ under $T'$, where $T_{1}$ follows $T_{0}$. Invalidates are batched and sent at the end of $T_{0}$. We assume that the backend can track keys that have been invalidated. We elaborate on this assumption in \S\ref{realizing-systems-design}. This means that if a key $k$ has been invalidated before the next write arrives at the backend, the backend does not need to send a second invalidate. 

Let the probability that a key has been invalidated at the end of an interval be $p$. Under invalidation policy, the \freshness is: $\frac{T'}{T}((1-p) \times P_W(T) \times c_i + p \times P_R(T) \times c_m)$.
The first term is the expected overhead of an invalidate at the end of $T_0$ (probability of the key not being invalidated multiplied by the probability of a write multiplied by the cost of an invalidate). The second term is the expected overhead of a miss over $T_1$ (probability of the key being invalidated multiplied by the probability of a read and multiplied by the cost of a miss). To calculate $p$, if the key has been invalidated in $T_0$ and there is a read, the key will be brought into the cache. if the key has not been invalidated in $T_0$ and there is no write, the key will not be invalidated in $T_1$. Hence:  $p = p P_R(T) + (1 - p)(1 - P_W(T))$. Solving: $
p = \frac{P_W(T)}{P_R(T) + P_W(T)}$.
If we substitute $p$ into $C_F$ and simplify, we get:
$
C_F =  \frac{T'}{T}\frac{P_R(T)P_W(T)}{ P_R(T) + P_W(T)} (c_m + c_i)
$. 
\staleness is $\frac{T'}{T}\frac{P_R(T)P_W(T)}{ P_R(T) + P_W(T)}$.

We now compare these costs with those for TTL-expiry calculated in \S\ref{sec:background-ttl}. 
We notice that \staleness for invalidates is strictly lower than \staleness for TTL-expiry: $\frac{T'}{T}P_R(T)$. 
Additionally, we note that for workloads that require real-time freshness, \freshness of invalidation is also lower than \freshness for TTL-expiry. 
For example, assuming request arrival is Poisson with $\lambda = 1$ and $r = 0.9$ and $T' = T$, \freshness of invalidation is $0.00892(c_i + c_m)$ and \freshness of TTL-expiry evaluates to $0.086c_m$, with the latter being significantly higher.
However, if workloads consist of mostly writes and not many reads, TTL-miss might be cheaper than invalidation; such scenarios are unlikely as caches are useful for workloads that have reads. 
 
In summary, reacting to writes enables real-time data freshness at lower overheads than TTL-based policies since invalidates and updates incur lower overheads than TTL-expiry and TTL-polling, respectively. 
However, we notice that invalidation is not \textit{strictly} better than update or vice versa based on \freshness and \staleness. 
This raises an interesting question: When should the data store update and when should it invalidate?  
We provide an initial answer to the question in the following sections.

\subsection{Picking between updates and invalidates}
\label{adaptive-algorithm}

The key challenge in picking between updates and invalidates is that their relative costs not only depend on the values of the system parameters (e.g., $c_m$ vs $c_u$) but also depend on the relative prioritization of the latency and throughput overheads (\staleness vs \freshness).
While the answer is clear in simple scenarios --- for instance, if one cares only about minimizing the latency (no matter the throughput cost), one would always send out updates, since they have $\staleness=0$ --- it is less clear in more complex scenarios. 
We now seek to answer this question for two such scenarios---when one seeks to maximize throughput irrespective of latency cost, and when one seeks to maximize throughput for a given latency cost (e.g., an SLO).

\myparagraph{Updates vs Invalidates when optimizing throughput.} We now describe a simple formula that decides whether to send the cache an update or invalidate upon receiving a write request to minimize the throughput overhead of freshness. 
We formulate the problem of deriving this formula in the style of classic online algorithms~\cite{fiat1998online}: we denote the gap between our online algorithm and the omniscient policy as $G$. The goal is for our policy to minimize $G$. Let $k$ be the probability of an update. $1-k$ is the probability of an invalidate. $k=1$ indicates that the policy decides to always update. Again let $T_0$ and $T_1$ be two consecutive intervals. Assume invalidates or updates are batched and sent at the end of $T_0$, and read and write are independent. We have: 
\begin{itemize}
    \item \textbf{Interval $T_1$ has at least a read} (probability: $P_R(T)$). The optimal decision is to do an update with cost $c_u$. 
    \item \textbf{Interval $T_1$ has no read but has at least a write.} (probability: $(1 - P_R(T)) P_W(T)$). The optimal decision is to do nothing with cost $0$. 
    \item \textbf{If interval $T_1$ has neither read nor write, consider $T_1$ skipped.} The intervals (say $T_2$) following $T_1$ will incur the same \textit{expected} gap $G$, down-weighted with probability $(1 - P_R(T))(1 - P_W(T))$. 
\end{itemize}
Each component of $G$ is the probability of an action ($k$ or $1-k$), times the probability of each of the three cases, and the cost difference to the optimal. Therefore, $G = (1-k)P_R(T)(c_i + c_m - c_u) + k (1-P_R(T)) P_W(T)c_u + (1-k) (1-P_R(T))P_W(T)  c_i + (1 - P_R(T))(1 - P_W(T)) G$.  $G$ is minimized when the coefficient of  $k$ is negative: 
$c_u < \frac{P_R(T)}{P_R(T) + P_W(T)} (c_m + c_i)$.
Intuitively, the policy should update if the cost of an update $c_u$ is lower than the cost of an invalidate (right-hand side). If $T \rightarrow 0$, the above formula reduces to 
$c_u < r (c_m + c_i)$.
This result is surprisingly simple since it tells us that whether to update or invalidate depends only on the read/write ratio of requests to an object. It is \emph{independent} of request rate $\lambda$ and $T$ when $T \rightarrow 0$. At small timescales ($T$ comparable to network delay), invalidates or updates have to be sent out immediately. Hence the decision should be independent of the exact values of $T$ and $\lambda$.

\begin{figure*}[t!]
\centering
\begin{subfigure}{.24\linewidth}
    \centering
 \includegraphics[width=\linewidth]{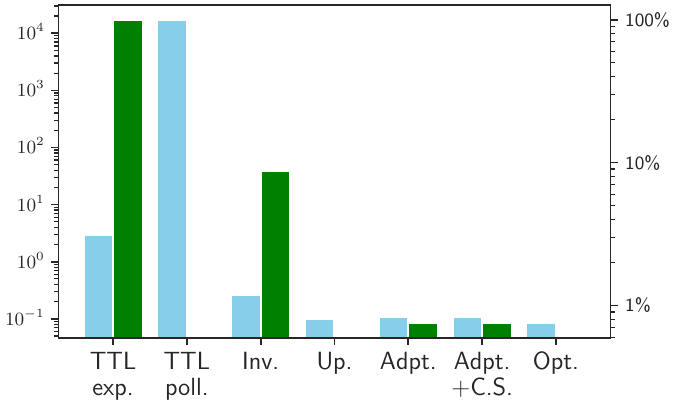}
    \caption{Poisson}
\end{subfigure}
~
\begin{subfigure}{.24\linewidth}
    \centering
 \includegraphics[width=\linewidth]{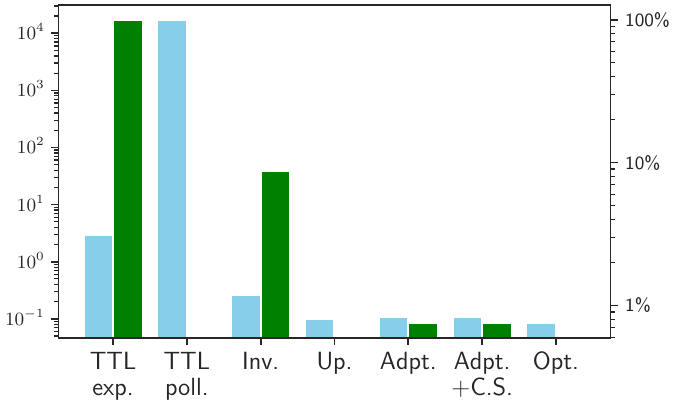}
    \caption{Poisson (Mix)}
\end{subfigure}
~
\begin{subfigure}{.24\linewidth}
    \centering
    \includegraphics[width=\linewidth]{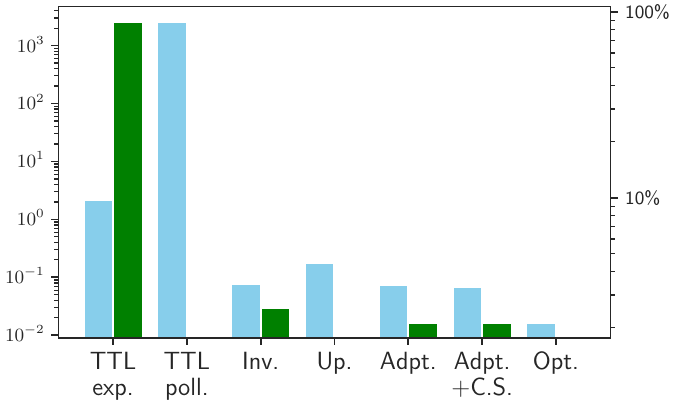}
    \caption{Meta
    }
\end{subfigure}
~
\begin{subfigure}{.24\linewidth}
    \centering
 \includegraphics[width=\linewidth]{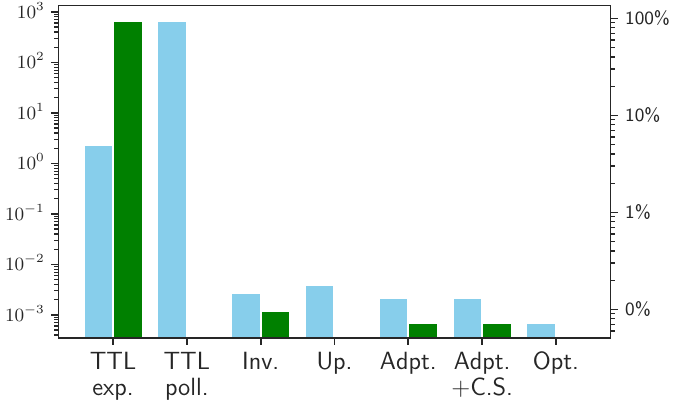}
    \caption{Twitter}
\end{subfigure}

\definecolor{grassgreen}{RGB}{83, 169, 83}  
\definecolor{skyblue}{RGB}{135, 206, 235}   

\caption{Comparison to baselines. Adpt. denotes our proposed adaptive policy. Adpt. + C.S. denotes data store knowing which keys are present in the cache  (C.S.). Opt. denotes the optimal policy. The left x-axis is in the log scale. The \textcolor{skyblue}{blue bar} indicates \textcolor{skyblue}{$C_F'$} (left axis, in $\times$), the \textcolor{grassgreen}{green bar} indicates \textcolor{grassgreen}{$C_S'$} (right axis, in $\%$), defined in \S\ref{sec:background-ttl}. The y-axis is in the log scale.}
\label{simulation-comparison}
\end{figure*}

\myparagraph{Maximizing throughput for a latency SLO.}
System designers rarely optimize throughput in isolation; instead, they typically seek to maximize throughput while meeting a latency target (e.g., an SLO). 
To address such scenarios, we extend the formulation we just described with an additional constraint to respect a given latency SLO. Since latencies are functions of implementations, we instead use \staleness (which represents the misses in the cache due to stale data) as a proxy for latency.  Thus, we seek to minimize the throughput overhead (\freshness) while meeting an upper-bound on the miss ratio due to staleness.

The staleness cost $C_S = \frac{T'}{T}\frac{P_R(T) P_W(T)}{P_R(T) + P_W(T)}$, the coefficient of $c_m$ in the formula for $C_F$. 
The miss ratio due to reading stale data ($C_S$ divided by the total number of reads $N_r$ over $T'$, or under Poisson, $\lambda rT'$) $C_S'$ (first introduced in \S\ref{sec:background-ttl}) is:
$
C_S' = \frac{1}{\lambda r T}\frac{P_R(T) P_W(T)}{P_R(T) + P_W(T)}
$.
If $T \rightarrow 0$, $C_S'$ reduces to: 
$1-r$. Staleness cost can be applied as a constraint from the user.
So, if $C$ is the user-specified $C_S'$ constraint: $C_S' \leq C$, the backend chooses to send updates if $(c_i + c_m) \times r > c_u$ or $1-r > C$, and chooses to send invalidates if otherwise. 
Once again surprisingly, we see that the choice is \emph{independent} of $\lambda$ and $T$ when $T \rightarrow 0$.
 
\subsection{Realizing the policy in a system}
\label{realizing-systems-design}

So far the discussion has been over parameters ($c_u$, $c_i$, and $c_m$) -- we next discuss preliminary ideas on how we can measure them in practice. Real workloads are often diverse and variable~\cite{twitter-cache}; invalidates or updates likely work in some situations but fall short in others. Therefore, these parameters need to be set adaptively in response to system bottlenecks and different overheads of invalidates and updates per key. 

\myparagraph{Estimating $c_u$, $c_i$, $c_m$ from systems bottlenecks.}

To estimate $c_u$, $c_i$, $c_m$, the policy first detects system bottlenecks that may arise from various components such as backend CPUs, caches, and network bandwidth. The policy then decides the cost values to set given the system bottleneck.

We have developed tools to identify systems bottlenecks, such as by measuring backend CPU utilization from \texttt{/proc/stat}, network usage from \texttt{/proc/net/dev}, and disk I/O usage from \texttt{/proc/diskstats}.
Users can also label a resource as the bottleneck based on offline profiling, which is often required before deployment.
The optimal strategy depends on the nature of the bottleneck. For instance, 
if the backend CPU or the network bandwidth is the bottleneck, $c_u$, $c_i$, and $c_m$ should be set based on either the CPU cycles needed for serialization, or message size. $c_u$, $c_i$ and $c_m$ should be scaled by the sizes of the actual keys and values. \autoref{tab:costs} illustrates one example of setting $c_u$, $c_i$, and $c_m$ where either the cache or backend CPU is the bottleneck. 
In scenarios where the user prioritizes read latency over throughput or always overprovisions, the policy can set $c_m = \infty$ and only send updates. 

\begin{table}[t!]
\centering
\begin{tabular}{lll}
\toprule
\textbf{Parameters} & \textbf{Breakdown} \\
\midrule
$c_m$: Miss & \begin{tabular}[c]{@{}l@{}}Cache: ser(K) + deser(K+V) + update \\ Data Store: deser(K) + read + ser(K+V) \end{tabular} \\
\midrule
$c_i$: Invalidation & \begin{tabular}[c]{@{}l@{}} Cache: deser(K) + delete \\ Data Store: ser(K) \end{tabular}  \\ 
\midrule
$c_u$: Update & \begin{tabular}[c]{@{}l@{}} Cache: deser(K+V) + update  \\ Data Store: ser(K+V) \end{tabular} \\  
\bottomrule
\end{tabular}
\vspace{1em}
\caption{An example of $c_u$, $c_i$, and $c_m$ where either the compute at the cache or the backend is the bottleneck. \texttt{ser} and \texttt{deser} refer to serialization and deserialization respectively. }
\vspace{-2em}
\label{tab:costs}
\end{table}

\myparagraph{Approximation with $E[W]$, the expected number of writes between reads.} From \S\ref{adaptive-algorithm}, while the overhead of invalidate and update depends on $P_R(T)$ and $P_R(T)$, We further introduce a pragmatic formula where we assume $T \rightarrow 0$. 
The formula in \S\ref{adaptive-algorithm} can be approximated: we measure $E[W]$, the expected number of writes between reads, and pick invalidate if $E[W] c_u < c_m + c_i$, and update otherwise. To explain, 
consider a sequence of writes followed by a read. To ensure that the read retrieves the freshest data, an update policy needs to send $E[W]$ number of updates, while an invalidation-based policy only needs to send the first invalidate (by tracking previously invalidated keys), skip sending invalidates for subsequent writes, and incur a miss upon the read. Tracking invalidated keys is feasible because keys are much smaller in size compared to values. The backend can also just track hot keys or recent invalidations. This can be done by simply maintaining a hashmap or storing an extra field in the database. The decision to invalidate or update depends on the relative overhead of the two approaches as decided in the formula. 

\myparagraph{Estimating $E[W]$ per-key with sketches}. We now discuss how $E[W]$ can be estimated per key. Exact $E[W]$ tracking requires three counters per key: $C_1$ stores the sum of $E[W]$ samples, and $C_2$ stores the number of $E[W]$ samples. $C_3$ stores the number of consecutive writes since the last read. To calculate the average $E[W]$, we divide $C_1$ by $C_2$. 
Upon write, we increment $C_3$. Upon read after a write, we add $C_3$ to $C_1$ and increment $C_2$ by 1. However, the overhead of exact tracking increases linearly with the number of keys and could become prohibitively expensive in practice.

One can lower storage overhead by estimating $E[W]$ with Count-min sketch~\cite{cormode2005improved}, which approximates read and write counters per key with a 2-D array. $E[W]$ can be estimated by dividing the number of writes by the number of reads. Upon reading or writing, the key passes multiple hash functions (one for each row) and is hashed into different columns of each row. To approximate the count for a key, we similarly hash the key multiple times and calculate the minimum of counters read from different columns that the key is hashed into. However, when the number of keys increases, one might obtain false positives due to hash collision.

To improve accuracy, we propose a modified Top-K sketch for better approximating the number of reads and writes. We keep the \textit{exact count} for Top-K most accessed keys while using the Count-min sketch to \textit{approximate} the count for the rest of the keys. This ensures that we get precise tracking for hot keys. A key can be promoted from Count-min sketch to Top-K if it becomes hot, or demoted from Top-K to Count-min sketch if it becomes cold. 

\begin{figure*}[t!]
\centering
\begin{subfigure}{.3\linewidth}
    \centering
 \includegraphics[width=\linewidth]{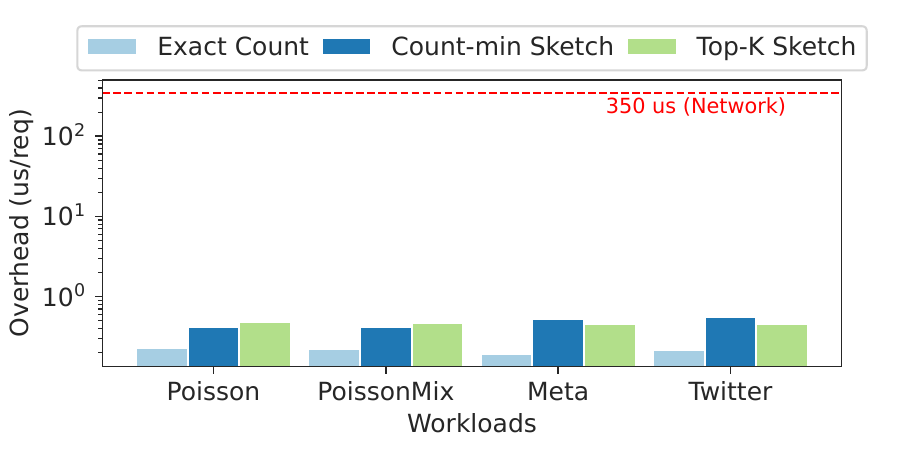}
    \caption{Latency Overhead}
\end{subfigure}
~
\begin{subfigure}{.3\linewidth}
    \centering
 \includegraphics[width=\linewidth]{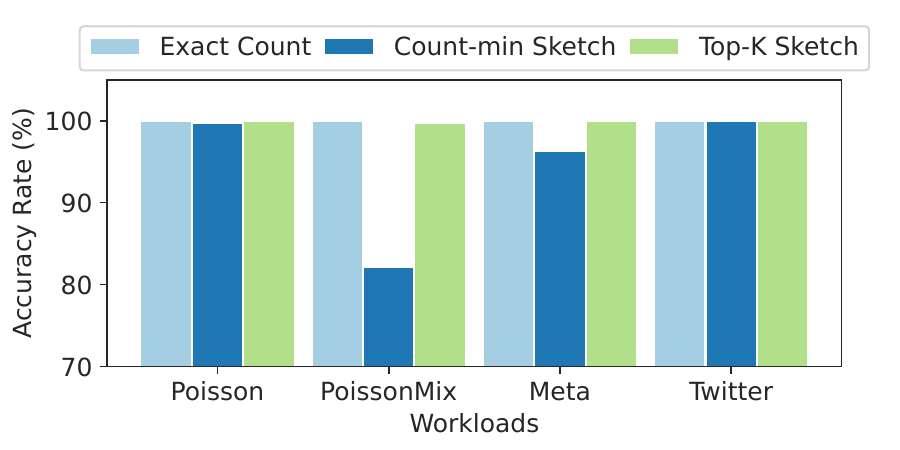}
    \caption{Prediction Accuracy}
\end{subfigure}
~
\begin{subfigure}{.3\linewidth}
    \centering
    \includegraphics[width=\linewidth]{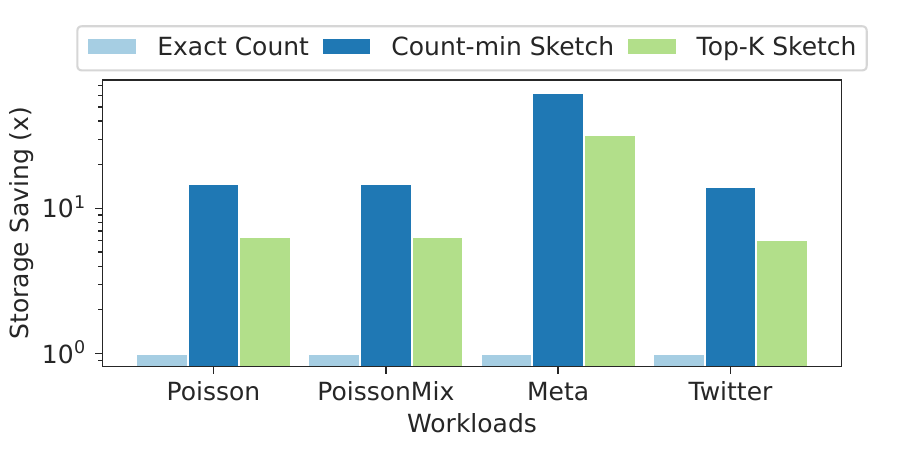}
    \caption{Storage Saving}
\end{subfigure}
\caption{Comparison of Latency, Prediction Accuracy, and Storage Saving across multiple sketches.}
\label{fig:sketches-comparison}
\end{figure*}

%% file: sections/related_works.tex
\subsection{Evaluation}

\myparagraph{How well does our policy perform?}
\label{optimal}
To evaluate our policy (Adpt.), we repeat the simulations performed in \S\ref{sec:background-ttl} with only throughput as the objective. 
We compare our policy against $6$ baselines: TTL-expiry, TTL-polling, always-update (Up.), always-invalidate (Inv.), along with 2 hypothetical policies. 
Adpt.+C.S assumes that the data store has knowledge of which keys are present in the cache; this enables it to send updates and invalidates only to relevant data objects. Comparing Adpt. with Adpt.+C.S once again evaluates our assumption about being able to evaluate freshness for different keys individually and additively (\S\ref{sec:background-model}).
The second hypothetical policy (Opt.) is an omniscient policy that has complete knowledge of both the cache contents and future requests and is optimal. 

We evaluate our policy on $4$ workloads; 3 from \S\ref{sec:background-model}, and a fourth that contains a 50-50 mix of two Poisson workloads, one that is read-heavy and another that is write-heavy. These workloads occur when sharing a cache across multiple applications, as is common practice today~\cite{cidon2017memshare}. 
Figure~\ref{simulation-comparison} presents the results. 
We draw three conclusions: (1)~reacting to writes provides significantly lower overheads than TTL-based policies, 
(2)~our policy equals or outperforms naive update and invalidation-based policies. 
(3)~while knowing the cache state can improve the overhead, our assumption about treating individual objects as independent and taking freshness decisions on a per-object basis is largely justified. 

\myparagraph{How well do sketches approximate $E[W]$ while lowering overhead?}
Overhead and accuracy of various sketches are presented in \autoref{fig:sketches-comparison}. Importantly, sketches do not need to determine the precise value of $E[W]$; they only need to decide whether $E[W] c_u < c_i + c_m$ (so it tolerates some inaccuracies in $E[W]$ estimation). We draw three observations: (1) The overhead of looking up $E[W]$ and maintaining the sketches for Top-K sketch and Count-min Sketch is negligible compared to the network delay. (2) Top-K sketch leads to good accuracy in deciding whether to invalidate or update. Count-min sketch can sometimes make wrong predictions. (3) Count-min sketch leads to the largest space saving followed by Top-K sketch. We suggest using the Top-K sketch to track $E[W]$ as it has high accuracy with significant space savings.

\section{Related Work}

TTL has been widely used~\cite{cachelib, twitter-cache, segcache, sultan2024ttls, faascache, modeling-ttl, perf-eval-ttl, exact-ttl-analysis, adaptive-ttl, ttl-cloud-caches, cost-aware-ttl, yang2023gl} and studied for in-memory caches: cache eviction~\cite{segcache}, estimating MRC~\cite{sultan2024ttls}, modeling hit ratio~\cite{jung2003modeling}, adaptive TTL for content delivery~\cite{basu2018adaptive}. 
However, prior works do not specifically target data freshness, or provide a framework for understanding the overhead of maintaining freshness. 

Cache invalidation has been explored: Meta~\cite{nishtala2013scaling} relies on invalidation to bound staleness at a larger timescale. MuCache~\cite{zhang2024mucache} explores cache invalidation for microservice graphs without blocking other accesses. Recent blog~\cite{cache-made-consistent} from Meta documents its system Polaris that detects and monitors inconsistencies introduced with cache invalidation. 
\cite{labrinidis2003balancing, yu1999scalable} explores whether to invalidate or materialize cached views for web caches. However, to the best of our knowledge, none of the prior works explore a quantitative model of data freshness based on bounded staleness; or an adaptive algorithm deciding between invalidation and update to maintain data freshness with a staleness bound $T$. 

%% file: sections/design_realization.tex
\section{Conclusion and Open Questions}
\label{sec: system-realization}

In this paper, we conclude that the path to efficient real-time cache freshness is by reacting to writes using updates and invalidates. While we developed a theoretical model to show the potential of such an approach, several key questions remain: 

\myparagraph{Ensuring guaranteed delivery of updates and invalidates.}
For TTL, data is guaranteed to expire after a specified time. 
However, lost or re-ordered updates and invalidates may cause a cached object to remain in a stale state in the cache indefinitely~\cite{cache-made-consistent}. 
This problem becomes more challenging in distributed and replicated caches since messages now have to be reliably multi-cast to the target caches. In the presence of resharding or node failures, ownership of keys can change. How to ensure that invalidation or update is propagated to the nodes that own the keys is itself a challenge. 

\myparagraph{Extending freshness formulation to many-to-many caching relationship.} Our algorithm (\S\ref{adaptive-algorithm}) assumes that one cached object can be mapped to one data store object.  
While this covers many workloads, some cached objects come from multiple reads from the backend data store. 
For example, the client can cache a web page, which requires rendering multiple data objects from the backend data store, such as figures, HTML fragments, and tables.
We believe we can extend our algorithm: a cached object has bounded staleness if its constituent parts satisfy the staleness bound, and \freshness and \staleness depend on the dependencies of the data read and written. 

\myparagraph{Combining freshness with eviction decisions.} 
We believe renewed attention to cache freshness will also uncover interesting questions on how to factor freshness decisions into cache eviction algorithms. 
While prior works have explored leveraging TTLs in eviction~\cite{segcache}, it is unclear how invalidation and updates can be co-designed with eviction since eviction algorithms can monitor the current value of the TTL timer, but cannot know when an invalidation or update is likely to arrive.